\documentclass[aps,twocolumn,pra,showpacs]{revtex4}
\usepackage[colorlinks,citecolor=blue,linkcolor=blue,hyperindex]{hyperref}
\usepackage{amsmath,mathrsfs,amsbsy,color,graphicx,bm,amsthm,amsfonts}
\usepackage{units,bbm,times,dcolumn,amssymb,epsfig,float}
\usepackage{color}
\begin{document}
\title{Quantum speed limit for a mixed initial state}
\author{Shao-xiong Wu$^1$\footnote{sxwu@nuc.edu.cn}}
\author{Chang-shui Yu$^2$\footnote{quaninformation@sina.com}}
\affiliation{$^1$ School of Science, North University of China,
Taiyuan 030051, China\\
$^2$ School of Physics, Dalian University of Technology, Dalian 116024, China}
\date{\today }
\begin{abstract}
A unified bound on the quantum speed limit is obtained for open quantum systems with the mixed initial state by utilizing the function of relative purity proposed in [Phys. Rev. Lett. 120, 060409 (2018)]. As applications, it is found that  the quantum speed limit bound for the damped Jaynes-Cumming model is determined by the competition among the non-Markovianity, the population of initial excited state and the initial-state coherence, which shows that  the population of initial excited state and the coherence of initial state can sharp the quantum speed limit despite that the non-Markovian effects can accelerate the evolution of open quantum system. For the dephasing model, a simple factorization law with the initial-state coherence shows that the quantum speed limit is only governed by the competition between the non-Markovianity and the coherence of the initial state.

\end{abstract}

\maketitle

\section{Introduction}
The quantum speed limit (QSL) is defined as the maximal evolution velocity  (the minimal evolution time) of the quantum system. This notion was first introduced by Anandan-Aharonov using the Fubini-Study metric \cite{Anandan90}. It plays significant role in the fields of quantum computation \cite{Lloyd00}, quantum metrology \cite{Giovannetti06PRL} and so on. In the closed system, the quantum speed limit describing  the minimal time of a quantum state evolving to its orthogonal state is given by the variation of energy  $\Delta E$ or the average of energy $\langle E\rangle$, that is the so-called Mandelstan-Tamm (MT) quantum speed limit is given by $\tau_{\text{qsl}}=\pi\hbar/(2\Delta E)$, and the Margolus-Levitin (ML) quantum speed limit is given by $\tau_{\text{qsl}}=\pi\hbar/(2\langle E\rangle)$ \cite{Mandelstam45,Fleming73,Bhattacharyya83,Bekenstein81,Vaidman92,Margolus98}. In the practical scenario, it is inevitable that the quantum system interacts with its surroundings, which has to be considered as the open quantum system \cite{Breuer02}.

In the recent years, the quantum speed limit in open systems have attracted extensive interest.
In \cite{Taddei13}, Taddei \emph{et al}.  investigated
the quantum speed limit in open systems by the bound of quantum Fisher information for the total Hilbert space of the system and its environment, which was developed in \cite{Escher11}. Using the relative purity, del Campo \emph{et al}. obtained the quantum speed limit which is analogous to the MT bound when the evolution of the open system is of Lindblad form \cite{Plenio13}. When the initial state is pure, Deffner and Lutz used the Bures angle to obtain the unified bound which covers the ML and MT quantum speed limit bound for the open systems, and showed that the non-Markovian effects could speed up the quantum evolution \cite{Deffner13}. Besides, many valuable efforts have also been devoted to some other aspects of the quantum speed limit in the open system such as  the initial-state dependence of the quantum speed limit \cite{Wu15}, the longitudinal and transverse type of quantum speedup \cite{Xu16}, the decoherence speed limit in the spin-deformed boson model \cite{Dehdashti15}, the classical-driving-assisted quantum speed-up \cite{Zhang15}, the role of the excited state population \cite{Xu14}, the quantum speed limit using experimentally realizable metric \cite{Pati16}, or using an alternative fidelity \cite{Ektesabi17}, or based on relative purity \cite{Zhang14}, or excluding rotating wave approximation \cite{Sun15} and the quantum speed limit in non-equilibrium environment \cite{Cai17}, or for multipartite open systems \cite{Liu15}.

In this paper, we use the function of relative purity \cite{Campaioli18} to develop a unified bound on the quantum speed limit (QSL) in open quantum systems. Using the von Neumann inequality and Cauchy-Schwarz inequality for operators, we obtain a unified bound on the QSL for the system with the mixed initial state. Based on this bound, we investigate the QSL in the damped Jaynes-Cumming model and dephasing model. For the damped Jaynes-Cumming model, the bound on the QSL is determined by the competition among the non-Markovianity, the initial excited state population, and the coherence of initial state. In \cite{Deffner13}, the non-Markovian effects can speed up the evolution of the quantum system, however, we find that the population of initial excited state and the coherence of initial state can sharp the bound on the QSL. In the dephasing model, the QSL is only governed by the non-Markovianity and the coherence of initial state, and is independent of the population of initial excited state. In particular, a simple factorization law with the population of initial state is found for the dephasing model. When the initial state is reduced to pure state, the bounds will become identical to the case in Ref. \cite{Wu15} for both models.
This paper is organized as follows. In Sec. II, we give the definition of quantum speed limit using the distance based on the  function of relative purity. In Sec. III and IV, we apply the definition of QSL to the  damped Jaynes-Cumming model and the dephasing model, respectively. The discussion and the conclusion are given in Sec. V.

\section{Quantum Speed limit for initial mixed state}
As a measure of statistical distance between quantum states, the Bures angle is defined as $\mathcal{L}(\rho,\sigma)=\arccos[F(\rho,\sigma)]$, where $F(\rho,\sigma)=\text{tr}[\sqrt{\sqrt{\rho}\sigma\sqrt{\rho}}]$ is the Uhlmann fidelity. In \cite{Deffner13}, the Bures angle was used as a distance to measure the quantum speed limit in the open quantum systems. Due to the poor computability, they only investigated the condition of the pure initial state, where the Bures angle was simplified as $\mathcal{L}(\rho_0,\rho_t)=\arccos[\langle\psi_0\vert\rho_t\vert\psi_0\rangle]$. In addition, the authors obtained the unified bound on the QSL and found the non-Markovianity can accelerate the evolution of quantum system.

In this paper, we will investigate the QSL for the mixed initial state. Here the distance we will use is the function of relative purity introduced in \cite{Campaioli18} where a tight QSL was provided for almost all the states in the unitary evolution, i.e., \begin{eqnarray}
\Theta(\rho_{0},\rho_t)=\arccos
\left(\sqrt{\frac{\text{tr}[\rho_{0}\rho_t]}{\text{tr}[\rho_{0}^{2}]}}\right).\label{eq:theta}
\end{eqnarray}
To give  the bound on the QSL, we first consider the time derivative of $\Theta$ as
\begin{align}
\frac{\text{d}}{\textrm{d}t}\Theta(\rho_{0},\rho_{t}) & \leqslant\left|\frac{\textrm{d}}{\textrm{d}t}\Theta(\rho_{0},\rho_{t})\right|\notag\\
&=\frac{\left|\frac{\text{tr}[\rho_{0}\dot{\rho_{t}}]}{\text{tr}[\rho_{0}^{2}]}\right|}
{2\sqrt{\frac{\text{tr}[\rho_{0}\rho_{t}]}{\text{tr}[\rho_0^{2}]}}
\sqrt{1-\frac{\text{tr}[\rho_{0}\rho_{t}]}{\text{tr}[\rho_{0}^{2}]}}}.\label{eq:tuidao}
 \end{align}
Substituting the definition (\ref{eq:theta}) into the Eq. (\ref{eq:tuidao}), one can obtain
\begin{eqnarray}
2\cos[\Theta]\sin[\Theta]\dot{\Theta}\textrm{tr}[\rho_{0}^{2}]\leqslant
\left|\textrm{tr}[\rho_{0}\dot{\rho_{t}}]\right|. \label{eq:tuidao3}
\end{eqnarray}
For the open systems, the time-dependent non-unitary evolution can be given by
\begin{eqnarray}
\dot{\rho}_{t}=L_{t}(\rho_{t}),\label{eq:Lrhot}
\end{eqnarray}
so, the Eq. (\ref{eq:tuidao3}) can be rewritten as
\begin{eqnarray}
2\cos[\Theta]\sin[\Theta]\dot{\Theta}\text{tr}[\rho_{0}^{2}]\leqslant\left|\text{tr}[\rho_{0}L_{t}(\rho_{t})]\right|.\label{eq5}
\end{eqnarray}

Based on the von Neumann inequality for operators
\begin{equation}
\left\vert\text{tr}[B_{1}B_{2}]\right\vert\leqslant
\sum_{i=1}^{n}\sigma_{1,i}\sigma_{2,i}
\end{equation}
with the descending singular values $\sigma_{1,1}\geqslant\cdots\geqslant\sigma_{1,n}$ and $\sigma_{2,1}\geqslant\cdots\geqslant\sigma_{2,n}$ for any complex $n\times n$ matrices $B_{1}$ and $B_{2}$,  one can have $\left|\text{tr}[\rho_{0}L_{t}(\rho_{t})]\right|\leqslant\sum_i p_i\lambda_i$, where $p_i$ are the singular values of the initial state $\rho_0$, and $\lambda_i$ are the singular values of  the operator $L_t(\rho_t)$. Since $p_i\leqslant1$, it is easy to obtain that $\sum_i p_i\lambda_i\leqslant\lambda_1\leqslant\sum_i\lambda_i$. The $\lambda_1$ is the largest singular value for the operator $L_t(\rho_t)$, which can be expressed as $\|L_t(\rho_t)\|_{\text{op}}$ and the sum of $\lambda_i$ can be expressed as the trace norm for the operator $L_t(\rho_t)$, i.e., $\|L_t(\rho_t)\|_{\text{tr}}=\sum_i\lambda_i$.  So, we can obtain the following  Margolus-Levitin type QSL bound
\begin{eqnarray}
\tau_{\text{qsl}}=\max\left\{ \frac{1}{\Lambda_{\tau}^{\text{op}}},\frac{1}{\Lambda_{\tau}^{\text{tr}}}\right\} \sin^{2}[\Theta(\rho_{0},\rho_{\tau})]\text{tr}[\rho_{0}^{2}],\label{qslml}
\end{eqnarray}
where the denominator in the above equation is defined as
\begin{eqnarray*}
\Lambda_{\tau}^{\text{op}}=\frac{1}{\tau}\int_{0}^{\tau}\text{d}t\| L_{t}(\rho_{t})\|_{\text{op}},~\Lambda_{\tau}^{\text{tr}}=\frac{1}{\tau}\int_{0}^{\tau}\text{d}t\| L_{t}(\rho_{t})\|_{\text{tr}}.
\end{eqnarray*}

Based on the Cauchy-Schwarz inequality for operators $\left\vert\text{tr}[B_1B_2]\right\vert^{2}\leqslant
\text{tr}[B_{1}^{\dagger}B_{1}]\text{tr}[B_{2}^{\dagger}B_{2}]$, the Eq. (\ref{eq5}) can be rewritten as
\begin{align}
&2\cos[\Theta]\sin[\Theta]\dot{\Theta}\text{tr}[\rho_{0}^{2}]\notag\\
\leqslant & \sqrt{\text{tr}[L_{t}(\rho_{t})L_{t}^{\dagger}(\rho_{t})]\text{tr}[\rho_{0}^{2}]}\notag\\
\leqslant &\sqrt{\text{tr}[L_{t}(\rho_{t})L_{t}^{\dagger}(\rho_{t})]}=\| L_{t}(\rho_{t})\|_{\text{hs}}.
\end{align}
The Hilbert-Schmidt norm is defined as $\|B\|_{\text{hs}} =\sqrt{\sum_{i}\sigma_{i}^{2}}$ with the singular value $\sigma_i$ for $n\times n$ matrix $B$. The second $``\leqslant"$ comes from the fact that the purity of quantum state satisfies $\text{tr}[\rho_{0}^{2}]\leqslant1$. So, the Mandelstam-Tamm type bound on the QSL for non-unitary dynamics (\ref{eq:Lrhot}) is
\begin{eqnarray}
\tau_{\text{qsl}}=\frac{1}{\Lambda_{\tau}^{\text{hs}}}\sin^{2}[\Theta(\rho_{0},\rho_{\tau})]\text{tr}[\rho_{0}^{2}],\label{qslmt}
\end{eqnarray}
where
\begin{eqnarray*}
\Lambda_{\tau}^{\text{hs}}=\frac{1}{\tau}\int_{0}^{\tau}\text{d}t\| L_{t}(\rho_{t})\|_{\text{hs}}.
\end{eqnarray*}
Combining the Eqs. (\ref{qslml}) and (\ref{qslmt}), the unified expression of the QSL bound for the mixed initial state is given by
\begin{eqnarray}
\tau_{\text{qsl}}=\max\left\{ \frac{1}{\Lambda_{\tau}^{\text{op}}},\frac{1}{\Lambda_{\tau}^{\text{tr}}},\frac{1}{\Lambda_{\tau}^{\text{hs}}}\right\} \sin^{2}[\Theta(\rho_{0},\rho_{\tau})]\text{tr}[\rho_{0}^{2}].
\end{eqnarray}

For the mentioned norms, we have the following inequality for the matrix $B$ \cite{Horn85},
\begin{eqnarray}
\|B\|_{\text{tr}}\geqslant\|B\|_{\text{hs}}\geqslant\|B\|_{\text{op}}.
\end{eqnarray}
So, it can lead to the following order of ``velocity" of quantum evolution $\Lambda_{\tau}^{\text{op}}\leqslant\Lambda_{\tau}^{\text{hs}}\leqslant\Lambda_{\tau}^{\text{tr}}$. Obviously, the ML-type bound based on the operator norm provides the sharpest QSL bound for open quantum systems. As applications, two exactly solvable examples to demonstrate the quantum speed limit for the initial mixed state are given in the latter part.

\section{The quantum speed limit for the damped Jaynes-Cumming model}

\begin{figure*}[t!]
  \centering
  \includegraphics[width=0.75\columnwidth]{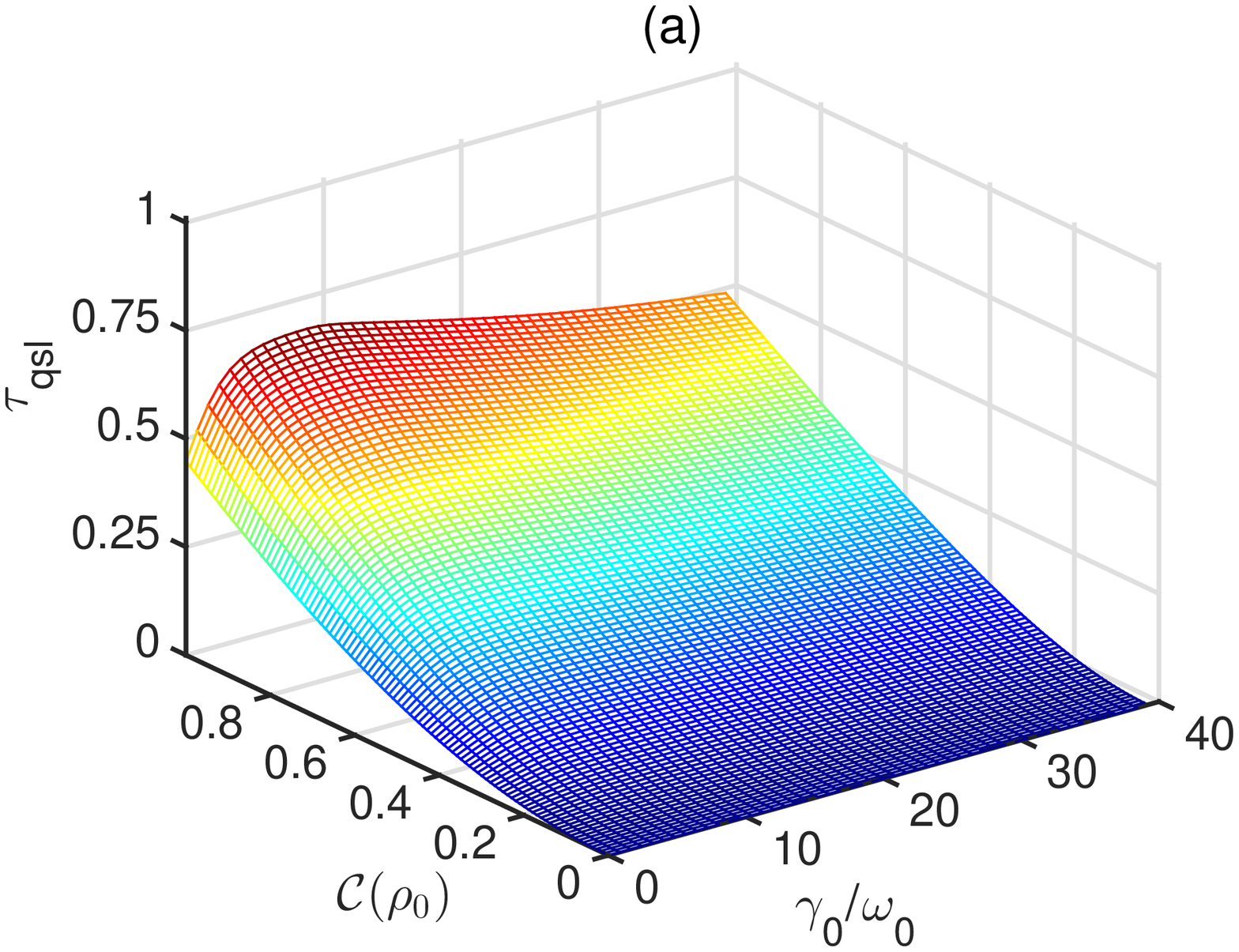}
    \includegraphics[width=0.75\columnwidth]{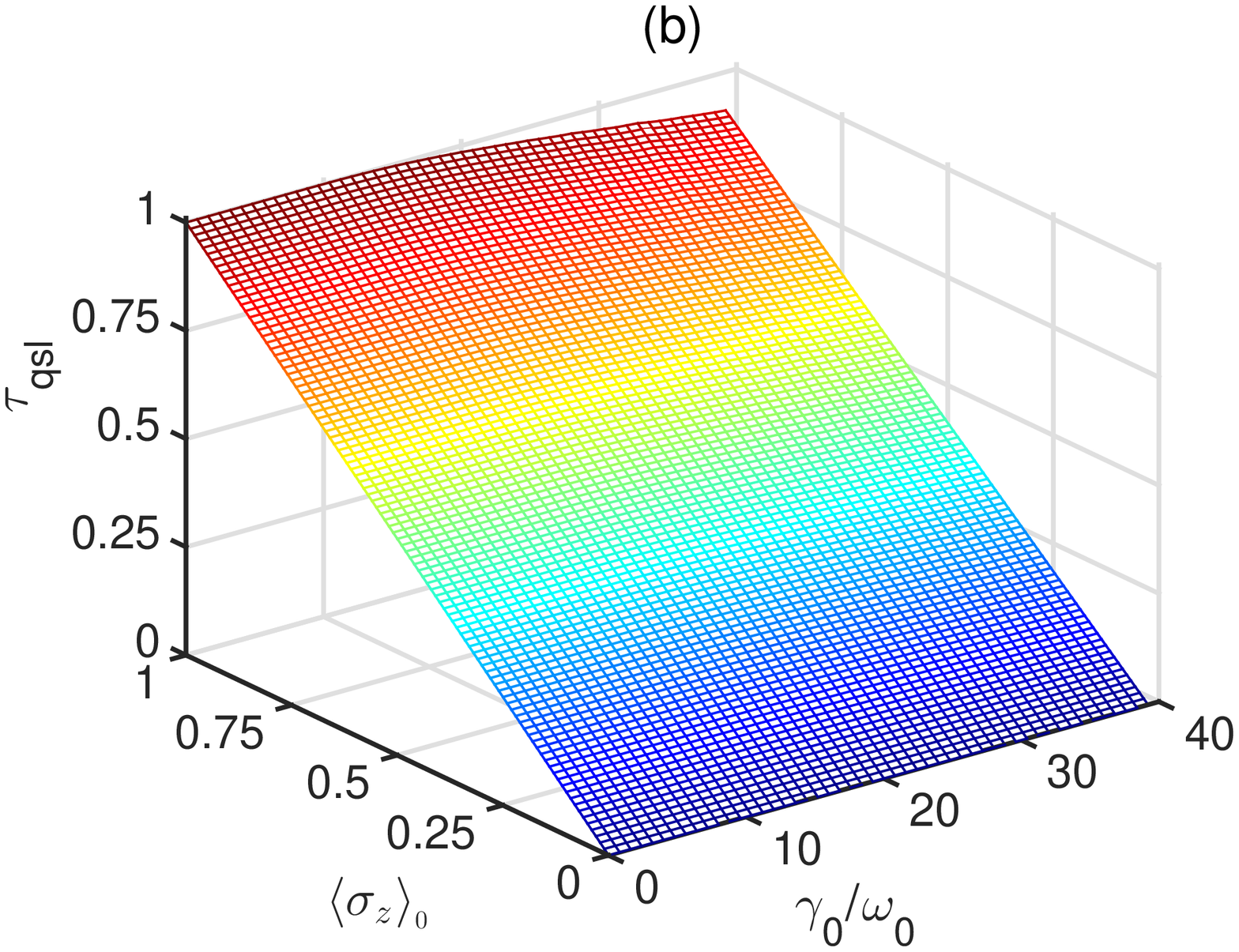}\\
        \includegraphics[width=0.75\columnwidth]{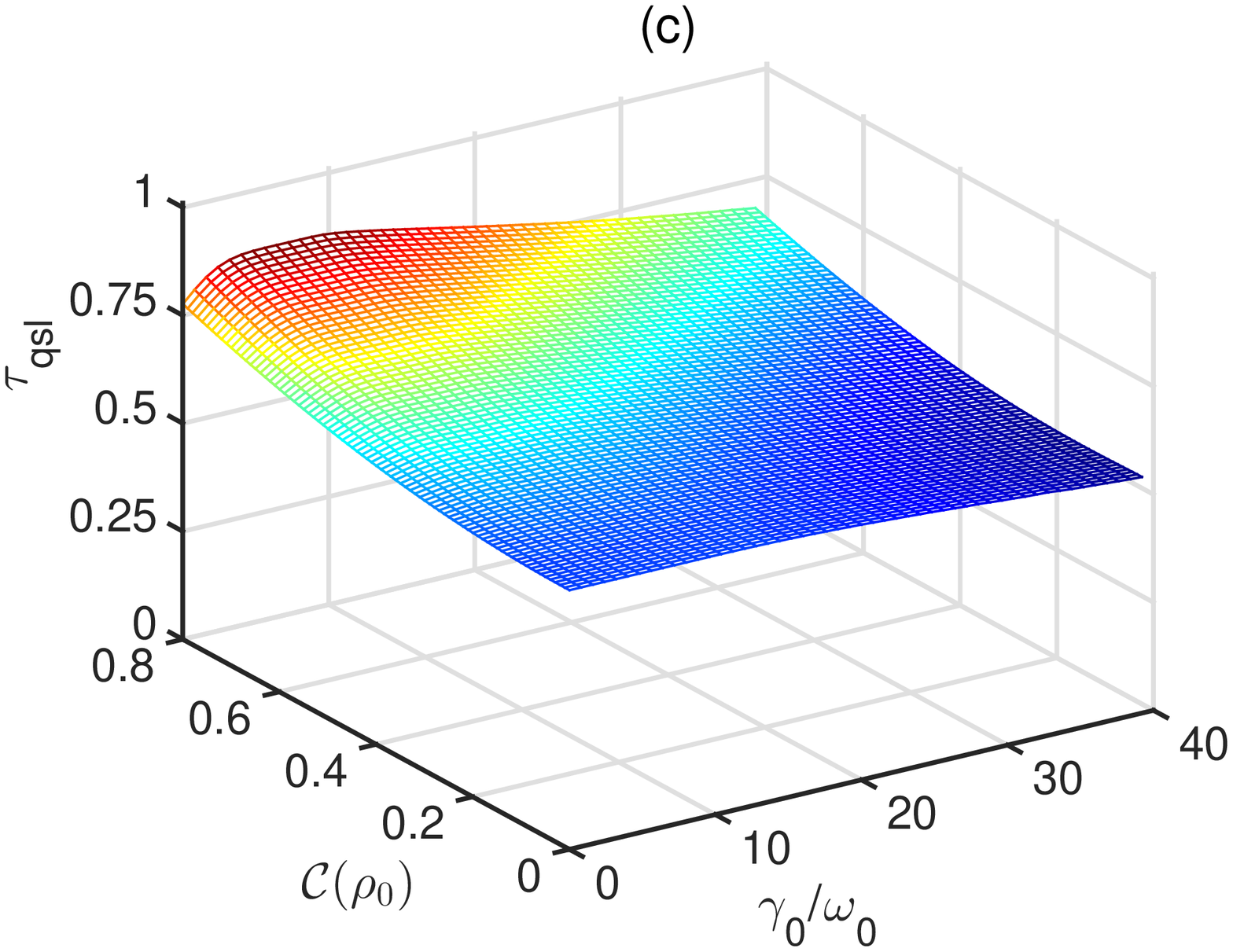}
        \includegraphics[width=0.75\columnwidth]{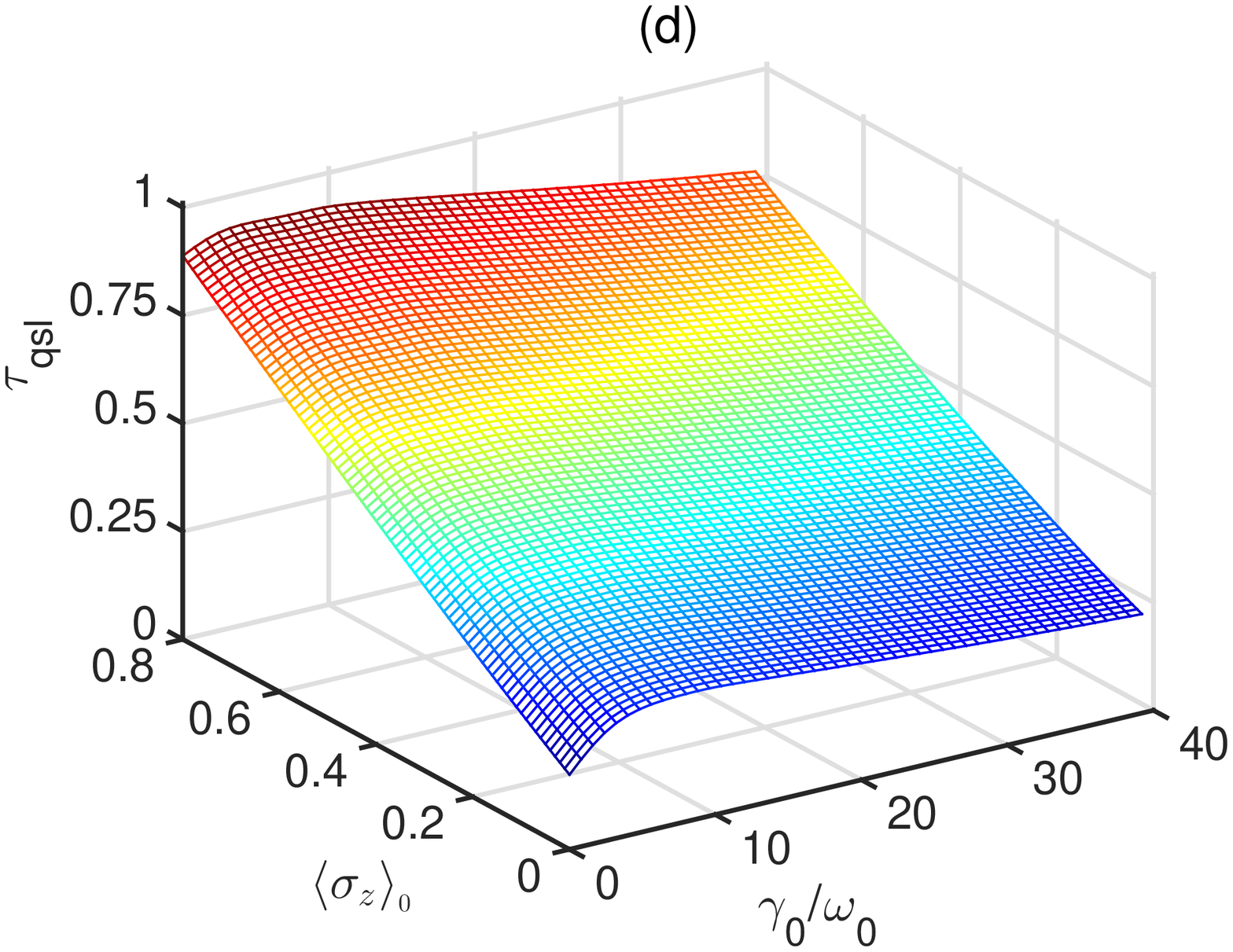}
  \caption{(Color online) The quantum speed limit bound $\tau_{\text{qsl}}$ for damped Jaynes-Cumming model. Panels (a) and (c), the variation of $\tau_{\text{qsl}}$ along with the coherence $\mathcal{C}(\rho_0)$ and coupling strength $\gamma_0$. The population of initial state is chosen as $\langle\sigma_z\rangle_{_0}=0$ in (a) and $\langle\sigma_z\rangle_{_0}=0.6$ in (c). Panels (b) and (d), the dynamics of $\tau_{\text{qsl}}$ along with the population of initial excited state $\langle\sigma_z\rangle_{_0}$ and coupling strength $\gamma_0$. The coherence of initial state $\mathcal{C}(\rho_0)$ is 0 in (b) and 0.6 in (d). The spectral width parameter is $\lambda=15$ (in unit of $\omega_0$) in the all panels. The actual driving time $\tau=1$.
  } \label{Fig1}
\end{figure*}

The damped Jaynes-Cumming model can be depicted as follows. The whole Hamiltonian of the system and reservoir is
\begin{eqnarray}
H=\frac{\omega_{0}}{2}\sigma_{z}+\sum_{k}\omega_{k}b_{k}^{\dagger}b_{k}+\sum_{k}\left(g_{k}\sigma_{+}b_{k}+\text{h.c}\right),
\end{eqnarray}
and the non-unitary dynamics of the reduced system can be described as
\begin{eqnarray}
L_{t}(\rho_{t})=\frac{\gamma_{t}}{2}\left(2\sigma_{-}\rho_{t}\sigma_{+}-\sigma_{+}\sigma_{-}\rho_{t}-\rho_{t}\sigma_+\sigma_-\right).
\end{eqnarray}
In the Bloch representation, the initial state is expressed as
\begin{eqnarray}
\rho_{0}=\frac{1}{2}\left(\begin{array}{cc}
1+r_{z} & r_{x}-\text{i}r_{y}\\
r_{x}+\text{i}r_{y} & 1-r_{z}
\end{array}\right),\label{eq:jcrho0}
\end{eqnarray}
and the reduced system at time $\tau$ is given as
\begin{eqnarray}
\rho_{\tau}=\frac{1}{2}\left(\begin{array}{cc}
(1+r_{z})\vert q_\tau\vert^2 & (r_{x}-\text{i} r_{y})q_\tau\\
(r_{x}+\text{i} r_{y})q_\tau^* & 2-(1+r_{z})\vert q_\tau\vert^2
\end{array}\right)
\end{eqnarray}
with $q_\tau=e^{-\Gamma_{\tau}/2}$, $\Gamma_\tau=\int_{0}^{\tau}\text{d}t\gamma_t$.

Without loss of generality, we assume the structure of reservoir is the Lorentzian form
\begin{eqnarray}
J(\omega)=\frac{\gamma_{0}}{2\pi}\frac{\lambda^{2}}{(\omega_{0}-\omega)^{2}+\lambda^{2}},
\end{eqnarray}
where $\lambda$ is the spectral width of reservoir and $\gamma_0$ is the coupling strength between the system and reservoir. The time-dependent decay rate  $\gamma_{t}=\frac{2\gamma_{0}\lambda\sinh(h\cdot t/2)}{h\cosh(h\cdot t/2)+\lambda\sinh(h\cdot t/2)}$, so the parameter $q_{\tau}$ can be given analytically
\begin{eqnarray}
q_{\tau}=e^{-\frac{\lambda\cdot\tau}{2}}\left[\cosh\left(\frac{h\cdot \tau}{2}\right)+\frac{\lambda}{h}\sinh\left(\frac{h\cdot \tau}{2}\right)\right]
\end{eqnarray}
with $h=\sqrt{\lambda^{2}-2\gamma_{0}\lambda}$.

The ML-type quantum speed limit is the tightest bound
\begin{eqnarray}
\tau_{\text{qsl}}=\frac{1}{\Lambda_{\tau}^{\text{op}}}\sin^{2}[\Theta(\rho_{0},\rho_{\tau})]\text{tr}[\rho_{0}^{2}],
\end{eqnarray}
and can be re-expressed by the following form using the Bloch vectors:
\begin{eqnarray}
\tau_{\text{qsl}}=\frac{(1-q_\tau)[r_{x}^{2}+r_{y}^{2}+r_{z}(1+r_{z})(1+q_\tau)]}
{\frac{1}{\tau}\int_0^{\tau}\left\vert\dot{q_t}\sqrt{r_{x}^{2}+r_{y}^{2}+4q_t^{2}(1+r_{z})^2}\right\vert\text{d}t}.\label{eq:jcqslvn}
\end{eqnarray}

In order to show the effect of quantum coherence, we will have to consider an analytic quantifier of quantum coherence \cite{Baumgratz14,yu1,yu2,yu3}. Here we use the $l_1$-norm-based coherence measure as  \cite{Baumgratz14,yu1} 
\begin{eqnarray}
\mathcal{C}(\rho)=\sum_{i\neq j}\vert\rho_{ij}\vert. \label{definationofcoherence}
\end{eqnarray}
So the quantum coherence of initial state $\rho_0$ (\ref{eq:jcrho0}) can be expressed as $\mathcal{C}(\rho_0)=\sqrt{r_x^2+r_y^2}$ and the Bloch vector $r_z$ can be replaced by the population of initial excited state $\langle\sigma_z\rangle_{_0}$. So the quantum speed limit (\ref{eq:jcqslvn}) can be rewritten as
\begin{eqnarray}
\tau_{\text{qsl}}=\frac{(1-q_\tau)[\mathcal{C}^{2}(\rho_{0})+\langle\sigma_{z}\rangle_{_0}(1+\langle\sigma_{z}\rangle_{_0})(1+q_\tau)]}
{\frac{1}{\tau}\int_0^{\tau}\left\vert\dot{q_t}\sqrt{\mathcal{C}^{2}(\rho_{0})+4q_t^{2}(1+\langle\sigma_{z}\rangle_{_0})^2}\right\vert\text{d}t}.\label{eq:JCqsl}
\end{eqnarray}

In the dynamics of damped Jaynes-Cumming model, when the spectral width of reservoir $\lambda$ and the coupling strength between the system and reservoir $\gamma_0$ meet the condition $\lambda\ll\gamma_0$, the non-Markovian effect will strongly influence the dynamics of the open system \cite{Breuer02,Breuer09}. Besides the non-Markovianity, the population of initial excited state $\langle\sigma_z\rangle_{_0}$ and the coherence of initial state $\mathcal{C}(\rho_0)$ also have effects on the QSL bound (\ref{eq:JCqsl}), and the comprehensive competition among them determines whether the evolution of open system can be accelerated. In Fig. \ref{Fig1}, we show that the variation of the ML-type QSL bound $\tau_{\text{qsl}}$ for the damped Jaynes-Cumming model. The spectral width parameter is $\lambda=15$ (in unit of $\omega_0$) and the actual driving time is $\tau=1$.

In Panels (a) and (c), we plot the change of the QSL bound along with the coherence of initial state $\mathcal{C}(\rho_0)$ and the coupling strength parameter $\gamma_0$. The population of initial excited state is $\langle\sigma_z\rangle_{_0}=0$ in panel (a), while $\langle\sigma_z\rangle_{_0}=0.6$ in panel (c). One can find that the competition between the non-Markovianity and the quantum coherence determines the ``velocity" of the evolution, and the quantum speed limit will be tighter when the quantum coherence is larger. In other words, the quantum coherence $\mathcal{C}(\rho_0)$ can make the QSL bound sharply, while the non-Markavianity can accelerate the evolution of the open system. Just like the initial-state dependence of quantum speed limit \cite{Wu15}, it can also demonstrate the  phenomenon of acceleration in the Markovian regime, which corresponds to $\gamma_0<7.5$.

In Panels (b) and (d), the variation of the quantum speed limit $\tau_{\text{qsl}}$ are shown along with the population of initial excited state $\langle\sigma_z\rangle_{_0}$ and the coupling strength $\gamma_0$. The coherence of initial state is $\mathcal{C}(\rho_0)=0$ in panel (b), while $\mathcal{C}(\rho_0)=0.6$ in panel (d). One can find that the QSL bound will be tighten with the value of $\langle\sigma_z\rangle_{_0}$ increasing. The competition between the non-Markovianity and the population of initial excited state determines whether the evolution of open system can be accelerated. When the initial state is excited state, i.e., $\langle\sigma_z\rangle_{_0}=1$ and $\mathcal{C}(\rho_0)=0$, we can find that the quantum speed limit is only accelerated by the non-Markovianity, which can be found in \cite{Deffner13}.

\begin{figure}[b!]
  \centering
  \includegraphics[width=0.95\columnwidth]{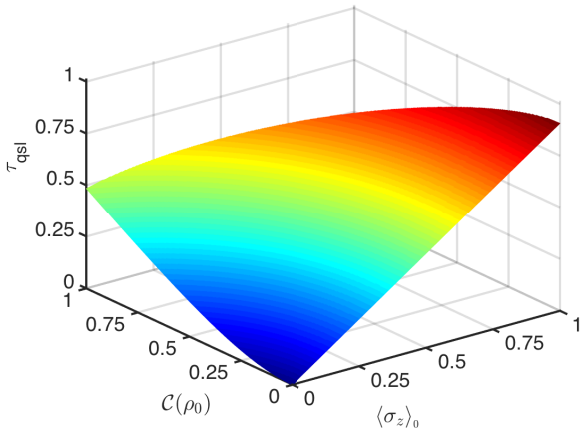}
  \caption{(Color online) The variation of quantum speed limit bound $\tau_{\text{qsl}}$ vs. the coherence of initial state $\mathcal{C}(\rho_0)$ and the population of initial excited state $\langle\sigma_z\rangle_{_0}$. The parameters are chosen as $\lambda=15$, and $\gamma_0=40$ (in unit of $\omega$). The actual driving time is $\tau=1$.
  } \label{Fig2}
\end{figure}

In Fig. \ref{Fig2}, we display that the variation of the quantum speed limit along with the quantum coherence of initial state $\mathcal{C}(\rho_0)$ and the population of initial excited state $\langle\sigma_z\rangle_{_0}$, where the non-Markovian parameters are chosen as $\lambda=15$ and $\gamma_0=40$ (in units of $\omega_0$) and the actual driving time is $\tau=1$. We can find that the competition between the coherence $\mathcal{C}(\rho_0)$ and population $\langle\sigma_z\rangle_{_0}$ determines whether the evolution can be accelerated. With the increasing of $\mathcal{C}(\rho_0)$ and $\langle\sigma_z\rangle_{_0}$, the QSL bound will be more sharp. One should notice that the quantum coherence of initial state $\mathcal{C}(\rho_0)$ and the population of initial excited state $\langle\sigma_z\rangle_{_0}$ should be satisfied the following inequality $\text{tr}[\rho_0^2]=\mathcal{C}^{2}(\rho_{0})+\langle\sigma_{z}\rangle_{_0}^2\leqslant1$, which is governed by the semi-positive property of the density matrix.

When the initial state is pure, the initial pure state can be written as $\vert\psi\rangle=\alpha e^{i\theta}\vert1\rangle+\sqrt{1-\alpha^{2}}\vert0\rangle$ (assuming the coefficient $\alpha\in\mathcal{R}$). Turning into the Bloch representation, we can obtain that $\mathcal{C}^2(\rho_0)=4\alpha^2(1-\alpha^{2})$, $\langle\sigma_z\rangle_{_0}=2\alpha^{2}-1$,
so the quantum speed limit can be simplified as
\begin{eqnarray}
\tau_{\text{qsl}}=\frac{\vert\alpha\vert(1-q_{\tau})[1-(1-2\alpha^{2})q_{\tau}]}
{\frac{1}{\tau}\int_0^{\tau}\left\vert\dot{q}_t\sqrt{1-(1-4q_t^2)\alpha^{2}}\right\vert\text{d}t},
\end{eqnarray}
which is matched with the Ref. \cite{Wu15}.

\section{The quantum speed limit for the dephasing model}

We will consider another exactly solvable model: the dephasing model. In the Schr\"{o}dinger picture, it can be expressed as a spin-boson type Hamiltonian describing a pure dephasing interaction between a qubit and a bosonic environment. The total Hamiltonian is
\begin{eqnarray}
H=\frac{\omega_0}{2}\sigma_{z}+\sum_{k}\omega_{k}b_{k}^{\dagger}b_{k}+\sum_{k}\sigma_{z}\left(g_{k}b_{k}^{\dagger}+g_{k}^{*}b_{k}\right),
\end{eqnarray}
and the dynamics of the reduced quantum system is
\begin{eqnarray}
L_t(\rho_{t})=\frac{\gamma_{t}}{2}\left(\sigma_{z}\rho_{t}\sigma_{z}-\rho_{t}\right).
\end{eqnarray}
In the Bloch representation, the initial state of system has the form (\ref{eq:jcrho0}), and the reduced system in time $\tau$ is
\begin{eqnarray}
\rho_{\tau}=\frac{1}{2}\left(\begin{array}{cc}
1+r_z & (r_x-\text{i}r_y)\text{e}^{-\Gamma_{\tau}}\\
(r_x+\text{i}r_y)\text{e}^{-\Gamma_{\tau}} & 1-r_z
\end{array}\right).
\end{eqnarray}
Taking the continuum limit of the bath mode and introducing the spectrum of reservoir is $J(\omega)$, one can find that the dephasing factor $\Gamma_{\tau}$ is given by \cite{Breuer02}
\begin{eqnarray}
\Gamma_\tau=\int_{0}^{\infty}\text{d}\omega J(\omega)\coth\left(\frac{\omega}{2k_{B}T}\right)\frac{1-\cos\omega \tau}{\omega^{2}},
\end{eqnarray}
where $k_B$ is the Boltzmann's constant and $T$ is temperature.

Without loss of generality, the environment spectrum can be chosen as the Ohmic-like spectrum with soft cutoff
\begin{eqnarray}
J(\omega)=\eta\frac{\omega^{s}}{\omega_{c}^{s-1}}\exp\left(-\omega/\omega_{c}\right),
\end{eqnarray}
where $\omega_{c}$ is the cutoff frequency and is assumed as unit, $\eta$ is the dimensionless coupling constant and the parameter $s>0$. The property of the environment is determined by the parameter $s$, and the reservoir is divided into the sub-Ohmic reservoir ($s<1$), Ohmic reservoir ($s=1$) and super-Ohmic reservoir ($s>1$). For the zero temperature condition, the dephasing factor $\Gamma_{\tau}$ can be solved analytically \cite{Chin13}
\begin{eqnarray}
\Gamma_\tau=\eta\left[1-\frac{\cos[(s-1)\arctan \tau]}{(1+\tau^{2})^{(s-1)/2}}\right]\Gamma(s-1),
\end{eqnarray}
where $\Gamma(\cdot)$ is the Euler gamma function.
Thus the ML-type QSL bound can be given by
\begin{eqnarray}
\tau_{\text{qsl}}=\frac{(1-\text{e}^{-\Gamma_{\tau}})\sqrt{r_{x}^{2}+r_{y}^{2}}}
{\frac{1}{\tau}\int_{0}^{\tau}\left|\gamma_{t}\text{e}^{-\Gamma_t}\right|\text{d}t}.\label{qsldephasing1}
\end{eqnarray}

For the zero temperature condition, the dephasing rate $\gamma_t$, i.e., the derivative of dephasing factor $\Gamma_t$, can be calculated as
\begin{align}
\gamma_t&=\int_0^{\infty}\text{d}\omega J(\omega)\frac{\sin \omega t}{\omega}\notag\\
&=\eta(1+t^2)^{-s/2}\Gamma(s)\sin[s\arctan t].
\end{align}
Following the concept of coherence (\ref{definationofcoherence}), the quantum speed limit $\tau_{\text{qsl}}$ (\ref{qsldephasing1}) can be rewritten as
\begin{eqnarray}
\tau_{\text{qsl}}=\frac{\mathcal{C}(\rho_0)(1-\text{e}^{-\Gamma_{\tau}})}
{\frac{1}{\tau}\int_{0}^{\tau}\left|\gamma_{t}\text{e}^{-\Gamma_{t}}\right|\text{d}t},\label{eq:deqsl}
\end{eqnarray}
which has a simple factorization law between the coherence of initial state $\mathcal{C}(\rho_0)$ and the quantum speed limit $\tau_{\text{qsl}}$. The competition between the coherence of initial state and the non-Markovianity determines whether the quantum evolution can be accelerated. One can find the fact that the QSL bound for the dephasing model (\ref{eq:deqsl}) is independent of the population of initial excited state $\langle\sigma_z\rangle_{_0}$, which is different from the damped Jaynes-Cumming model.

When the initial state is pure, such as $\vert\phi\rangle=\beta\text{e}^{i\theta}\vert1\rangle+\sqrt{1-\beta^2}\vert0\rangle$, the coherence of initial state is $\mathcal{C}(\rho_0)=2\beta\sqrt{1-\beta^2}$ and the quantum speed limit can be simplified as
\begin{eqnarray}
\tau_{\text{qsl}}=\frac{2\beta\sqrt{1-\beta^2}(1-\text{e}^{-\Gamma_{\tau}})}
{\frac{1}{\tau}\int_{0}^{\tau}\left|\gamma_{t}\text{e}^{-\Gamma_{t}}\right|\text{d}t},
\end{eqnarray}
which is consistent with the result in \cite{Wu15}.

\section{Discussion and conclusion}

We investigate the quantum speed limit for the open quantum systems with the mixed  initial state  based on the function of relative purity introduced in \cite{Campaioli18}. With the applications in the damped Jaynes-Cumming model and the dephasing model, we find that the QSL bound is determined by the competition among the non-Markovianity, the population of initial excited state $\langle\sigma_z\rangle_{_0}$, and the coherence of initial state  $\mathcal{C}(\rho_0)$ for the damped Jaynes-Cumming model. Even though the quantum evolution can be accelerated by the non-Markovian effects, the population of initial excited state and the coherence of initial state can make the quantum speed limit sharply. While, for the dephasing model, the quantum speed limit is only governed by the non-Markovianity and the coherence of initial state which is shown by a simple factorization law with the coherence of initial state. When the initial state is reduced to pure state, the quantum speed limit can be identical to Ref. \cite{Wu15} for both models. The Bures angle distance is the function of Uhlmann fidelity, the distance used in this paper is the function of relative purity. Even though the two measures are equal for pure state, the quantum speed limit for initial mixed state based on the Bures angle deserves endeavor in our further investigation.

\section*{ACKNOWLEDGMENT}
Wu was supported by the National Natural Science Foundation
of China under Grant No. 11747022, the Doctoral Startup Foundation of North
University of China (No. 130088), and the Science Foundation of North
University of China (No. 2017031). Yu was supported by the National Natural Science Foundation of China, under Grant No.11775040 and No. 11375036, and the Fundamental Research Fund for the Central
Universities under Grants No. DUT18LK45.

\end{document}